%Paper: hep-ph/9406223
%From: ejw@cuphyf.phys.columbia.edu (Erick J. Weinberg)
%Date: Fri, 3 Jun 1994 13:27:11 -0400
%Date (revised): Sun, 5 Jun 1994 23:28:01 -0400

\magnification=1200
\baselineskip=14truept
\def\pr#1#2#3#4{Phys.~Rev.~D{\bf #1}, #2 (19#3#4)}
\def\prl#1#2#3#4{Phys.~Rev.~Lett. {\bf #1}, #2 (19#3#4)}

\def\np#1#2#3#4{Nucl.~Phys.~{\bf B#1}, #2 (19#3#4)}
\def\pl#1#2#3#4{Phys. Lett. {\bf #1B}, #2 (19#3#4)}

\def\half{{ {1\over 2}}}
\def\tr{\mathop{\rm tr}}
\def\seff{S_{\rm eff}}
\def\veff{V_{\rm eff}}
\def\fv{{\rm fv}}
\def\b{{\rm b}}
\def\ct{{\rm ct}}

\def\vh{\hat V}
\def\zh{\hat Z}
\def\bphi{\bar\phi}

\def\sqr#1#2{{\vcenter{\hrule height.#2pt
        \hbox{\vrule width.#2pt height#1pt \kern#1pt
                \vrule width.#2pt}
        \hrule height.#2pt}}}
\def\square{\mathchoice\sqr64\sqr64\sqr{2.1}3\sqr{1.5}3}

% Make @ letter...
\catcode`@=11
\def\@versim#1#2{\lower.7\p@\vbox{\baselineskip\z@skip\lineskip-.5\p@
    \ialign{$\m@th#1\hfil##\hfil$\crcr#2\crcr\sim\crcr}}}
\def\simge{\mathrel{\mathpalette\@versim>}} %
\def\simle{\mathrel{\mathpalette\@versim<}} %
\catcode`@=12 % at signs are no longer letters

\line{\hfil CU-TP-633}
\line{\hfil hep-ph/9406223}
\vglue .5in

\vskip 1truein
\noindent{\bf BUBBLE NUCLEATION IN THEORIES WITH \hfill\break
SYMMETRY BREAKING BY RADIATIVE CORRECTIONS}
\vskip 42truept
\noindent \hskip 1truein \vbox{\hsize=4in\noindent Erick J. Weinberg \hfill
\vskip 14pt \noindent  Physics Department    \hfill \break
Columbia University  \hfill  \break
New York, NY 10027   \hfill  }
\vskip 42truept
\noindent{\bf INTRODUCTION}
\vskip 14pt
\topskip 0truept

\noindent\footnote{}{\noindent Based on a talk given at the NATO Advanced
Research Workshop Electroweak Physics and the Early Universe, Sintra,
Portugal, March 25, 1994 \hfill\break\noindent
 This work was supported in part by
the US Department of Energy}

    In any first order phase transition a quantity of fundamental
importance is the rate at which bubbles of the equilibrium (``true
vacuum") phase nucleate within the metastable (``false vacuum") phase.
The bubble nucleation rate per unit volume, $\Gamma$, can
be calculated by a method, due to Coleman\rlap,{$^1$} which is based on
finding a ``bounce" solution to the
classical Euclidean field equations.  Thus, for a theory with a
single scalar field, one seeks a solution of
$$  \partial_\mu\partial_\mu\phi \equiv \square \phi
    = {\partial V\over \partial\phi}
           \eqno(1)  $$
for which the $\phi$ is near its true vacuum value near the origin,
but approaches its false
vacuum value as any of the $x_\mu$ tend to $\pm\infty$.

     A difficulty arises when one deals with theories where the
symmetry breaking is a result of radiative corrections\rlap.$^2$  In such cases
the vacuum structure is not determined by $V(\phi)$, but instead can be
found only by examining the effective potential, $\veff(\phi)$.  The
bounce equation (1)  is clearly inappropriate --- in fact, if
$V(\phi)$ has only a single minimum there will not even be a bounce
solution.  An obvious solution is to replace $V(\phi)$ by
$\veff(\phi)$ in the bounce equation\rlap.$^3$

     Although plausible, and clearly a step in the right direction,
this procedure raises some questions.  The one-loop radiative
corrections generate an effective action which contains not only
$\veff$, but also terms involving derivatives (of all orders) of the
fields.  Can these terms be neglected when dealing with
configurations, such as the bounce solution, which are not constant in
(Euclidean) space-time?  Even if this can be done in a first
approximation, what are the nature and magnitude of the corrections
which these terms generate?  There are also questions relating to
$\veff$ itself.  First, the effective potential obtained by
perturbative calculations differs considerably from that defined by a
Legendre transform (the latter must be convex, while the former is
not).  The latter clearly does not lead to an appropriate bounce, but
how precisely does the formalism pick out the former?  Further, the
perturbative effective potential is known to be complex for certain
values of the fields.  How is the imaginary part of
$\veff$ to be handled?

     In this talk I will describe a systematic calculational
scheme$^4$ which gives an answer to these questions.  The general idea
is to use the path integral approach of Callan and Coleman\rlap,$^5$
but to integrate out certain fields at the outset.  This leads to a
modified effective action that gives a correct description of the
vacuum structure of the theory and has a bounce solution that can
provide the basis for a tunneling calculation.  To leading
approximation, one obtains the same result for $\Gamma$ as would have
been obtained by replacing $V$ by $\veff$ in the standard procedure.
The next-to-leading terms, which can be expressed in terms of the
effective action, give calculable and significant corrections.  The
corrections beyond these, although well-defined, do not have a simple
expression in terms of the effective action.  In particular, the
potentially complex terms in the effective potential do not appear
directly, but only as part of more complicated functional determinants
that can easily be shown to be real.

     Throughout this talk I will confine myself to the case of bubble
nucleation by quantum mechanical tunneling at zero temperature.
Similar issues arise in connection with finite temperature bubble
nucleation, even in theories where radiative corrections have
little effect on the zero-temperature vacuum structure.  It should be
possible to deal with these by appropriate extensions of the
methods described here.

\vskip 28truept
\noindent{\bf A SIMPLE MODEL}
\vskip 14truept

     Although radiative symmetry breaking is of greatest interest in
the context of gauge theories, I will begin by using a somewhat simpler
model to illustrate the method.  This avoids
some technical complications associated with gauge theories, which I
discuss later.  The model has two scalar fields, $\phi$ and $A$, and
is governed by the Lagrangian
$$    {\cal L} = {1\over 2} \left(\partial_\mu\phi\right)^2
       + {1\over 2} \left(\partial_\mu A\right)^2
        -V(\phi,A)
     \eqno(2)$$
with
$$  V = {1\over 2} \mu^2 \phi^2 + {\lambda\over 4!} \phi^4
    + {1\over 2} m^2 A^2  +  {f\over 4!} A^4
       + {1\over 2}g^2\phi^2 A^2  .
     \eqno(3) $$
Both $\mu^2$ and $m^2$ are positive, so the tree-level potential has
only a single symmetric minimum.   We would like $A$-loop corrections
to induce symmetry breaking with $\langle \phi \rangle = \sigma$ and
$\langle A \rangle = 0$.  This is done by choosing parameters
so that $\mu = O(g^2\sigma)$,  $m = O(g\sigma)$, $f= O(g^2)$, and
$\lambda=O(g^4)$.

     The one-particle-irreducible Green's functions of this theory are
generated by the effective action $\seff$, which can be expanded in the
familiar derivative expansion
$$ S_{\rm eff}(\phi,A) = \int d\,^4x \left[ V_{\rm eff}(\phi,A)
    + \half Z_\phi(\phi,A) (\partial_\mu\phi)^2
    + \half Z_A(\phi,A) (\partial_\mu A)^2
      +   \cdots \right]
           \eqno(4) $$
where the dots represent terms containing four or more derivatives.
The leading contribution to the effective potential is of order $g^4$.
It is given by the sum of the tree-level potential, all graphs
with a single $A$-loop, and the corresponding counterterms and is
equal to
$$ \eqalign {(V_{\rm eff})_{g^4}(\phi) &=
   {1\over 64\pi^2 }(m^2 + g^2 \phi^2)^2
   \left[ \ln  \left( {m^2 + g^2 \phi^2
  \over m^2 + g^2 \sigma^2 } \right) - {1\over 2} \right] \cr
    &\qquad   -{1\over 4}\left({\mu^2 \over \sigma^2} + y_1\right)
      \left( \phi^2 - \sigma^2\right)^2 }
      \eqno(5) $$
\topinsert\vskip1.9truein
\font\tenrm=cmr10 scaled 833
\font\tenbf=cmbx10 scaled 833
\noindent {\tenbf Figure 1. \tenrm The two-loop graphs which contribute
to the effective potential at order $g^6$.  Solid  and dashed lines
represent $\phi$ and $A$ propagators, respectively. }\endinsert
\vskip 14truept
\noindent where $y_1$ is a constant of order $g^2m^2/\sigma^2$.  The order
$g^6$ contributions to $V_{\rm eff}$ arise from the two graphs shown in
Fig.~1, together with several  counterterm graphs.  In these graphs
the propagators are not simply those of the
tree-level theory, but take into account the interaction with a
background $\phi$ field.  Thus, the $A$ propagator is given by
$$ G_A(k^2) = i (k^2  -m^2 - g^2 \phi^2)^{-1} .
     \eqno(6)$$
The $\phi$ propagator is a bit more complicated.  Because the
tree-level and one $A$-loop contributions are of the same magnitude,
the effects of both must be included, giving
$$ G_\phi(k^2) =
    i \left[k^2  -\mu^2 -(V_{\rm eff})_{g^4}'' \right ]^{-1} .
     \eqno(7)$$
(The inclusion of one-loop effects in this propagator means that when
calculating higher order corrections one must be careful to avoid
double-counting of graphs; this will not be relevant for the calculations
discussed here.) At order $g^8$ there are both three-loop
graphs containing $A$ propagators and graphs with a single
$\phi$ loop.  The latter graphs bring in terms
proportional to $\ln  (V_{\rm eff})_{g^4}''$ that become
complex for certain values of $\phi$.

     This complex effective potential deserves some comment.   The
effective potential is often defined formally through a Legendre
transform of the generating functional of connected Green's functions.
The properties of Legendre transforms then imply that it is everywhere
convex, a property that the perturbatively calculated effective
potential does not enjoy in theories with multiple stable or metastable
vacua.  To understand this, recall that the effective potential
$\veff^{\rm Leg}(\hat\phi)$  obtained by Legendre transform is equal to
the minimum value of the energy density among all states $|\Psi \rangle$
such that $\langle \Psi | \phi(x) | \Psi \rangle = \hat\phi $. For values
of $\hat\phi$ that lie between two vacua, the energy is minimized by
states $|\Psi \rangle$ which are superpositions of the two vacua; these
lead to a flat effective potential.  The perturbative
effective potential is most easily understood$^6$ by decomposing $\phi({\bf
x}, t)$ into a spatially uniform mode $\phi_0(t)$ and a part $\tilde
\phi({\bf x}, t)$ whose spatial integral vanishes.   In the infinite
volume limit the former mode is essentially classical, and one can
discuss states in which the wave functional is of the form
$\delta(\phi_0 -\hat\phi) \tilde \Psi [\tilde\psi({\bf x}]$.
The real part of the perturbative
effective potential is the minimum expectation value of the energy
density among states of this form subject to the additional condition that
$\tilde \Psi [\tilde\psi({\bf x}]$ be concentrated near $\tilde\psi=0$.
For values of $\hat\phi$ between two vacua these latter states are not
eigenfunctions of the Hamiltonian and eventually decay to states
whose wavefunctionals are concentrated on configurations with large
fluctuations in $ \tilde\psi({\bf x})$.  Being unstable, they have
complex energies  with the imaginary part of the energy related to their
decay rate.  In the region where the tree-level
potential is concave the decay is essentially classical and is
reflected in a perturbative imaginary part to the effective potential.
Beyond this region decay proceeds by quantum tunneling and leads a
nonperturbative imaginary part.

\vskip 28truept
\noindent{\bf THE CALLAN-COLEMAN FORMALISM}
\vskip 14truept

    In theories where the vacuum structure can be read off from the
tree-level Lagrangian, $\Gamma$ can be calculated using a formula that
was derived by  Callan and Coleman$^5$ using path integral techniques.
The starting point is the quantity
$$ \eqalign {G(T)&= \langle \phi({\bf x})=\phi_{\fv}|\,e^{-HT}\,|
   \phi({\bf x})=\phi_{\fv} \rangle \cr
           &=\int [d\phi]e^{-[S(\phi) +S_{\ct}(\phi)] }\cr}
    \eqno(8)  $$
where the Euclidean action
$$ S(\phi) =\int d\,^3x \int^{T/2}_{-T/2}dx_4
   \left[{1 \over 2} (\partial_\mu\phi)^2 + V(\phi)\right]
       \eqno(9) $$
and $S_{\ct}$ contains the counterterms needed to make the theory finite.

    The path integral is over all configurations such that
$\phi$ takes its false vacuum value $\phi_{\fv}$ at $x_4=\pm T/2$ and
at spatial infinity. In the limit $T \rightarrow \infty $,
Eq.~(8) is dominated by the lowest energy state with a
non-vanishing contribution (i.e., the false vacuum) and is of the
form
$$ G(T) \approx e^{-{\cal E} T\Omega}
   \eqno(10) $$
where $\Omega$ is the volume of space and $\cal E$ may be
interpreted as the energy density of the false vacuum state.  Because
this is an unstable state, $\cal E$ is complex with its
imaginary part giving the decay rate, which in this case is
simply the bubble nucleation rate.  Dividing by $\Omega$ gives
the nucleation rate per unit volume,
$$  \Gamma = -2 \,{\rm Im}\, {\cal E} .
    \eqno(11) $$

     The path integral may be approximated as the sum of the
contributions about all of the stationary (or quasi-stationary)
points of the Euclidean action $S(\phi)$: the pure false vacuum,
the bounce solution $\phi_b$ with all possible locations in Euclidean
space-time, and all multibounce configurations.
In each case the
contribution to the path integral is obtained by expanding the
field about the classical solution $\bphi(x)$:
$$ \phi(x) = \bphi(x) + \eta(x)
  \eqno(12)$$
and then integrating over $\eta$.  To leading approximation one
keeps only the terms in the action which are quadratic in $\eta$.
Expanding these in terms of the normal modes of $ S''(\bphi) =
-\square + V''(\bphi) $ gives a product of Gaussian integrals.  The
evaluation of these integrals is completely straightforward in the case of
the false vacuum, but about the bounce solution is complicated by the fact
that  $ S''(\phi_b)$ has four zero and one negative eigenvalues.  The zero
modes are treated by introducing collective coordinates; integrating
over these gives a factor of $\Omega T$.  The
negative mode is handled by deforming the contour of integration;
aside from a factor of $1/2$, this gives a contribution whose
imaginary part is just that which would have been obtained from a
naive application of the Gaussian integration formula.  Finally, the
contributions from the multibounce configurations are simply related to
that from the single bounce.   Summing over all of these stationary
points gives
$$  G = G_0 + \Omega TG_0G_b + \half (\Omega T)^2 G_0G_b^2  +\dots
  = G_0 e^{\Omega T G_b} .
  \eqno(13) $$
This leads to
$$  \Gamma= 2\,{\rm Im} G_b =    e^{- B} K J
    e^{-[S_{\ct}(\phi_\b) - S_{\ct}(\phi_{\fv})] }
    \, (1 + \cdots)
    \eqno(14) $$
where
$$ B= S(\phi_\b) - S(\phi_{\fv})
    \eqno(15)$$
and
$$  K =  \left|{ {\det}'[- \square + V''(\phi_\b)] \over
     {\det}[- \square + V''(\phi_{\fv})]} \right|^{-1/2} .
 \eqno(16)$$
Here $\det '$ indicates that the translational zero-frequency
modes are to be omitted when evaluating the determinant.  The
determinants are divergent, but these divergences are cancelled
by the terms containing $S_{\ct}$.
Finally, $J$ is a
Jacobean factor associated with the introduction of the collective
coordinates; it can be shown to equal  $B^2/4\pi^2$.

      Scaling arguments can be used to estimate the magnitude of the
various terms in the expression for $\Gamma$. Let us assume that we can
identify a small coupling  $\lambda$ and a dimensionful quantity $\sigma$
such that the potential can be written in the form
$$ V(\phi) = \lambda \sigma^4 U(\psi)
    \eqno(17) $$
where $U$ involves no small couplings and the dimensionless field
$ \psi = {\phi /\sigma} $.
The minima of the potential must then be located either at
$\phi=0$ or at values of $\phi$ of order $\sigma$.

     By defining a dimensionless variable $ s = \sqrt{\lambda}\,
\sigma x$, we may write the field equations as
$$    \square_s \psi = {\partial U\over \partial \psi}  .
      \eqno(18) $$
{}From the assumptions made above, this equation involves no small
parameters and so has a bounce solution in which $\psi$ is
of order unity and differs from the false vacuum within a
region of a spatial extent (measured in terms of
$s$) which is also of order unity.  In terms of the original
variables, the bounce has $\phi$ of order $\sigma$ and
extends over a range of $x$ of order $1/(\sqrt{\lambda}
\,\sigma)$.

    With the same change of variables, the action becomes
$$  S = {1 \over \lambda} \int d\,^4s\,
  \left[{1\over 2} \left({\partial\psi \over \partial s^\mu}\right)^2
         + U(\psi)\right]  .
    \eqno(19) $$
Since the integrand contains no small parameters, while the
volume of the bounce restricts the integration to a region of
order unity, the bounce action $B$ is of order $\lambda^{-1}$.

    Similarly, the determinant factor $K$
becomes
$$ \eqalign {K &= {1\over 2}
     \left|{ {\det}'[(\lambda\sigma^2) (- \square_s +
U''(\psi_\b))] \over
     {\det}[(\lambda\sigma^2) (- \square_s + U''(\psi_{\fv}))]}
      \right|^{-1/2} \cr
    &=  {1\over 2} \lambda^2\sigma^4
     \left|{ {\det}'[- \square_s + U''(\psi_\b)] \over
     {\det}[- \square_s + U''(\psi_{\fv})]} \right|^{-1/2} }
 \eqno(20)$$
where the explicit factor of $\lambda^2\sigma^4$ on the second line
arises because the ${\det}'$ factor involves four fewer modes than the
$\det$ factor.  With this factor extracted, the ratio of determinants
is formally of order unity, although divergent.
Finally, the
Jacobean factor $J$ is proportional to $B^2 \sim
\lambda^{-2}$.  Putting all of these factors together, we see
that the nucleation rate is of the form
$$  \Gamma = a \sigma^4\, e^{-b/\lambda}
    \eqno(21)$$
with $a$ and $b$ both of order unity.  In
practice $b$ can be calculated rather accurately, since it is not
difficult to numerically solve the differential equation for the
bounce solution.  On the other hand, an accurate determination of
$a$, which involves a functional determinant, is exceedingly
difficult.

\vskip 28truept
\noindent{\bf A MODIFIED FORMALISM}
\vskip 14truept

     If we were to apply this formalism to our scalar field model,
the first step would be to find a bounce solution to the Euclidean
field equations implied by the Lagrangian (2).  The problem is that,
because the tree level potential has only the symmetric minimum, there
is no bounce solution.  We can circumvent this problem by integrating
out the $A$ field from the start, and writing
$$  G(T)= \int [d\phi][dA] e^{-S(\phi,A)}
            \equiv \int [d\phi] e^{-W(\phi)} .
        \eqno(22)$$
$W$ may be thought of as a kind of effective action, although it is
not the same as the more usual $\seff$ of Eq.~(4).  Graphically, the two
actions differ in that $W$ receives contributions only from graphs with
only internal $A$-lines and external $\phi$-lines.  Also, $W$ is
divergent while $\seff$, since it generates renormalized Green's
functions, must be finite.  Performing the path integral over $A$ fields,
one obtains
$$  \eqalign{ W(\phi) &= \int d\,^4x \,
   \left[ {1\over 2} \left(\partial_\mu\phi\right)^2
      + {1 \over 2} \mu^2 \phi^2 + {\lambda \over 4} \phi^4
    + {1 \over 2} \langle x|
  \ln \left[ -\square + M^2(\phi)  \right] |x\rangle  \right]\cr
        & \qquad + {\rm counterterms} + \cdots }
       \eqno(23) $$
where $M^2(\phi) = m^2 + g^2\phi^2$ and the dots represent the
two-loop and higher order corrections.

     Although $W$, in contrast with the tree-level action, does
reflect the true vacuum structure of the theory, we cannot simply
proceed by solving its field equations to obtain a bounce.  There are two
difficulties here.  First, we only have a perturbative expansion
for $W$.  Second, and more importantly, $W$ is nonlocal and will therefore
lead to quite complicated field equations.

    Although nonlocal, $W$ can be approximated by a local functional if
$\phi$ is sufficiently slowly varying. (For the terms shown explicitly in
Eq.~(23), the requirement is that the change in $\phi$ over a distance of
order $M(\phi)^{-1}$ be small.)   This local functional takes the form of a
derivative expansion
$$ W(\phi) = \int d\,^4x \left[ \vh(\phi)
    + \half \zh(\phi) (\partial_\mu\phi)^2
      +   \cdots \right] .
           \eqno(24) $$
Although this is similar in form to the expansion  for $\seff$, the
functions entering the two expansions are not the same. In
line with the remarks made previously, $\vh(\phi)$ and $\zh(\phi)$ differ
from $V_{\rm eff}(\phi,0)$ and $Z_\phi(\phi,0)$ by the omission of graphs
with internal $\phi$-lines.  In particular, to $O(g^4)$ we have
$$ \vh_{g^4} = (\veff)_{g^4}
    \eqno(25) $$
but at the next order
$$ \vh_{g^6} \ne (\veff)_{g^6}
    \eqno(26) $$
because the second graph in Fig.~1 contributes to $\veff$ but not to
$\vh$.

     Returning to the bubble nucleation problem, let us define an action
$$  W_0(\phi) = \int d\,^4x \left[ {1\over 2} (\partial_\mu \phi)^2
   + \hat V_{g^4}(\phi) \right]
     \eqno(27) $$
that does display the correct vacuum structure and that, at least for
slowly varying $\phi$, is a good approximation to $W$.  We can now
attempt to evaluate the path integral over $\phi$ in Eq.~(22) by expanding
about the stationary points $\bphi$ of $W_0$.  These include the
homogeneous false vacuum and a bounce solution obeying
$$  \square \phi = {\partial \hat V_{g^4}\over \partial\phi} .
           \eqno(28)  $$
Defining $\eta(x) = \phi(x) - \bphi(x)$, we obtain
$$  W(\phi) = W(\bphi) + \int d\,^4z \, W'(\bphi;z) \eta(z)
     + {1\over 2} \int d\,^4z\, d\,^4z' \, W''(\bphi;z,z')\eta(z) \eta(z')
     + O(\eta^3)
     \eqno(29) $$
where primes denote variational derivatives.  Note that $W'(\bphi;z)$ does
not vanish, since $\bphi$ is a stationary point of $W_0$ but not of
$W$.  Note also that we cannot use the derivative expansion of $W$ inside
the path integral, since the integral includes rapidly varying field
configurations.
Inserting Eq.~(29) into the path integral and then proceeding as in the
standard case, we eventually obtain
$$  \Gamma = e^{-C_1} \, e^{C_2}
  \left| {{\det}' W''(\phi_\b) \over {\det} W''(\phi_{\fv}) }
      \right|^{-1/2} \, J  (1 + \cdots)
    \eqno(30) $$
where
$$  C_1 = W(\phi_\b) - W(\phi_{\fv})
  \eqno(31) $$
and
$$  C_2 = W'(\phi_\b) [W''(\phi_\b)]^{-1} W'(\phi_\b)
       - (\phi_\b \rightarrow \phi_\fv)  .
  \eqno(32)$$

    Let us now examine the various terms in this expression, beginning
with $C_1$.  Scaling arguments of the type described
previously show that the bounce solution to Eq.~(28) has a characteristic
spatial size of order $1/(g^2\sigma)$ and is therefore slowly varying
relative to the scale $M(\phi)$ entering the one-loop approximation to $W$,
Eq.~(23).  This both allows us to use the derivative expansion (24) to
evaluate $W(\phi_b)$ and also suppresses the higher-derivative
terms in this expansion.  Doing the derivative expansion, one finds that
the leading contribution to $C_1$ is
$$   \hat B_0 = \int d\,^4x \left[ {1\over 2} (\partial_\mu \phi_b)^2
   + \hat V_{g^4}(\phi_b)
    - (\phi_\b \rightarrow \phi_\fv)   \right]   \sim {1\over g^4} .
     \eqno(33) $$
Recalling that $\hat V_{g^4}= (\veff)_{g^4}$, we see that this is just the
result one would expect from simply replacing $V$ by $\veff$
in the standard formalism.
The next-to-leading contribution is
$$ \hat B_1 =  \int d\,^4x \left[ {1\over 2}\hat Z_{g^2}(\phi_b)
             (\partial_\mu \phi_b)^2
   + \hat V_{g^6}(\phi_b) - (\phi_\b \rightarrow \phi_\fv)
                   \right]   \sim {1\over g^2} .
     \eqno(34) $$
Because $\hat V_{g^6} \ne (\veff)_{g^6}$ (although $\hat Z_{g^2}$ and
$(Z_\phi)_{g^2}$ are equal in our model), this term is not simply the next
correction to $\seff$.  The term beyond this, which is of order unity,
involves the $O(g^8)$ contributions to $\hat V$ and the $O(g^4)$
contributions to $\hat Z$, as well as the leading four-derivative terms.

    For $C_2$ we need both $W'$ and $W''$.  The derivative expansion can
be used for the former;  because $\bphi$ is a stationary point of $W_0$,
the leading contribution here is from $\delta W' \equiv W'- W'_0 $.
Matters are less simple for $W''(\bphi;z,z')$. Although the
derivative expansion can be used when $|z-z'|$ is large, the
behavior for small $|z-z'|$ is sensitive to the high momentum
modes and so the derivative expansion fails no matter how slowly
varying $\bphi$ is.  However, the relation
$$  W'' = W_0'' \left[ 1 + (W_0'')^{-1} \delta W'' \right]
      \eqno(35)$$
can be used to obtain formal expansions for
$(W'')^{-1}$ and  $\det W''$ as power series in $(W_0'')^{-1} \delta
W''$.   The actual utility of these expansions depends on the size of
the contribution from the region of small $|z-z'|$. In the
calculation of $C_2$ this contribution is subdominant and  $(W'')^{-1}$
can be approximated  by $(W_0'')^{-1}$; one finds that $C_2$
is of order unity.  For the determinant factor, on the other hand, more
terms must be  retained:
$$ \eqalign{  {\det}'[W''] &= {\det}'[W_0'']
   \det [ I + (W_0'')^{-1} \delta W''] \cr
    &= {\det}'[W_0'']
    \exp\left\{\tr \ln [ I + (W_0'')^{-1} \delta W''] \right\} \cr
     &= {\det}'[W_0'']
    \exp\left\{\tr (W_0'')^{-1} \delta W''
      + {1\over 2} \tr \left[(W_0'')^{-1} \delta W''\right]^2
                + \cdots\right\} \,.}
        \eqno(36) $$
In the last line, the first term in the exponent is of order $1/g^2$
while the second is of order unity.  In fact, the former term is given by
the second graph of Fig.~1 which, it will be recalled, contributes to
$\veff$ but not to $\hat V$.  It  combines with the terms in $\hat
B_1$ to give the full $O(g^6)$ contribution to the  effective potential.
Combining all  other factors together in a prefactor $A$ of order
unity, we obtain
$$ \Gamma = A \sigma^4e^{-(B_0+B_1)}
      \eqno(37) $$
where
$$  B_0 =\hat B_0 =
    \int d\,^4x \left\{ \left[ (V_{\rm eff})_{g^4}(\phi_\b)
     +  {1\over 2}(\partial_\mu\phi_\b)^2 \right]
     -(\phi_\b \rightarrow \phi_{\fv}) \right\} \sim {1\over g^4}
      \eqno(38) $$
$$  B_1 = \int d\,^4x \left\{ \left[ (V_{\rm eff})_{g^6}(\phi_\b)
     + {1\over 2}( Z_\phi)_{g^2}(\phi_\b) (\partial_\mu\phi_\b)^2 \right]
     -(\phi_\b \rightarrow \phi_{\fv}) \right\} \sim {1\over g^2} .
      \eqno(39) $$

    After seeing how the expansion of the determinant  combines
with the exponent factors  to reconstruct $\seff$  to
leading and next-to-leading orders, it is natural to speculate that this
process might continue to higher orders.  It does not.  In particular, it
is at the next order that we would encounter the potentially complex
contributions to $\veff$ which arise from graphs with a single
$\phi$ loop.  These would appear in a derivative expansion of the
determinant factor
$$  \left| {{\det}' W_0''(\phi_\b) \over {\det} W_0''(\phi_{\fv}) }
      \right|^{-1/2}  .
    \eqno(40) $$
However, a derivative expansion of $\det W_0''$ is valid only for fields
which vary slowly relative to $(\veff)''_{g^4}$. The bounce
solution does not satisfy this condition, and so the complex terms in
$\veff$ do not appear.  (One can make a derivative expansion of the
determinant in the denominator, but $(\veff)_{g^8}(\phi_{\fv})$ is real.)

\vskip 28truept
\noindent{\bf SCALAR QUANTUM ELECTRODYNAMICS}
\vskip 14truept

These methods can be applied to gauge theories with only minor
modifications.   To be specific, consider the case of scalar
electrodynamics with the quartic scalar self-coupling taken to be
$O(e^4)$ so that the vacuum structure is determined by the
one-loop corrections to the effective potential\rlap.$^2$  Directly
following the
approach used for the scalar field example, one would integrate out the
photon field at the start to obtain an effective action for the complex
scalar field.  The classical equations following from this action have a
bounce solution of the desired type which gives the leading approximation
to $\Gamma$.  However, the calculation of the next order terms turns out
to be much more complicated than in the scalar case.  These complications
can be avoided by an alternative approach.  Since the bounce solution can
be chosen to be entirely real, it is possible to integrate out both the
photon field and the imaginary part of the scalar field to give an action
$W(\phi)$ which depends only on a single real scalar field.   The analysis
then proceeds very much as before.  The $O(e^4)$ terms in $\veff$ and
$\hat V$ are identical and, together with the dominant gradient term,
lead to a contribution to the exponent proportional to $e^{-4}$.  At
order  $e^6$ there are two graphs
contributing to $\veff$, neither of which appears in $\hat V$.  Both are
recovered from the expansion of $\det W''$ and combine with
$(Z_\phi)_{e^2} = (\hat Z)_{e^2}$ to give a contribution to the exponent
which is $O(e^{-2})$.    $\veff$ becomes
complex at order $e^8$ but, as before, the failure of the derivative
expansion of $\det W_0''$ about the bounce solution prevents the
offending terms from entering the nucleation rate calculation.   In
fact, the appearance of infrared divergences at small $\phi$ means that
the derivative expansion cannot be carried out beyond the
four-derivative terms.  Instead, one must extract the potential and
two-derivative terms, leaving a remainder which gives an $O(1)$
contribution to $\Gamma$ that can be absorbed in the prefactor.

      The issue of gauge-dependence also arises.  The nucleation rate
is a measurable physical quantity and so should be independent of gauge.
However, the formulas above express it in terms of the effective
potential and other gauge-dependent$^{7-10}$ quantities.   The Nielsen
identities$^8$
indicate how this conflict can be resolved.  In a class of gauges
with  photon propagator
$$  D_{\mu\nu}(k^2) = {-g_{\mu\nu} - {k_\mu k_\nu\over k^2} \over k^2} +
     \xi {k_\mu k_\nu\over k^4}
    \eqno(41) $$
these identities take the form
$$  \xi {\partial \seff \over \partial\xi} = \int d\,^4x C[\phi(x)]
          {\delta \seff \over \delta\phi} \, .
     \eqno(42) $$
(We do not need the explicit form of the functional $C[\phi(x)]$, but only
the fact that it is $O(e^2)$.)  For a uniform $\phi$ field this reduces
to
$$  \xi {\partial \veff \over \partial\xi} = C(\phi) {\partial\veff
                     \over \partial\phi} \,.
       \eqno(43) $$
Now let us expand this in powers of $e^2$ to yield a series of
identities.  The first of these states that $(\veff)_{e^4} $ is
gauge-independent, as is easily verified.  The next is
$$   \xi {\partial (\veff)_{e^6} \over \partial\xi} = C_{e^2}(\phi)
     {\partial(\veff)_{e^4} \over \partial\phi} \, .
       \eqno(44) $$
Further identities are obtained by making derivative expansions of both
sides of Eq.~(42) and expanding the various terms in powers of $e^2$. The
first new identity obtained in this manner is
$$   \xi {\partial (Z_\phi)_{e^2} \over \partial\xi} =
           {\partial C_{e^2} \over \partial\phi}  \,.
       \eqno(45) $$

    Let us now apply these identities to $\Gamma$.  The leading
($O(1/e^4)$) term in the exponent involves $(\veff)_{e^4}$, and is
manifestly gauge-independent.  The next term in the exponent, of order
$1/e^2$, contains $(\veff)_{e^6}$ and $(Z_\phi)_{e^2}$.  Using the
identities we have just obtained, we find that the gauge dependence of
this term is given by
$$  \eqalign{ \xi {\partial\over \partial\xi }\int d\,^4x \left[
   (Z_\phi)_{e^2}(\partial_\mu\phi)^2 +  (\veff)_{e^6} \right] &=
     \int d\,^4x \left[ {\partial C_{e^2} \over \partial\phi}
       (\partial_\mu\phi)^2 +
       C_{e^2} {\partial(\veff)_{e^4} \over \partial\phi} \right]\cr
    &= \int d\,^4x \left[ (\partial_\mu C_{e^2} )(  \partial_\mu\phi)  +
       C_{e^2} {\partial(\veff)_{e^4} \over \partial\phi}\right] \cr
     &= \int d\,^4x  C_{e^2} \left[ - \square \phi
        + {\partial(\veff)_{e^4} \over \partial\phi}\right]\,. }
   \eqno(46) $$
Because $\phi_b$ is a solution of the Euclidean field equations, the last
line vanishes.

\vskip 28truept
\noindent{\bf CONCLUSION}
\vskip 14truept

    In this talk I have described how the decay rate of a metastable
vacuum can be calculated in a theory whose vacuum structure is
determined by radiative corrections.  As in the standard case, the
result may be written as a dimensionful prefactor times the
exponential of an action involving a bounce solution.  To leading
approximation this exponent is just the tree-level action supplemented
by the dominant one-loop contribution to the effective potential.  The
first correction to the exponent arises from the next-to-leading
contributions to the effective potential and the leading correction to
the tree-level kinetic part of the effective action.  Although smaller
than the leading terms, these give an addition to the exponent which
is larger than order unity and is thus more important than the
prefactor.  It does not appear that this correction need have any
particular sign, but rather that it might increase the nucleation rate
in some theories and reduce the rate in others.

   All further corrections may be absorbed into the prefactor.
Although some of these can be identified with particular terms in the
effective action, this is not true of all higher corrections.
Specifically, the graphs which give rise to complex terms
in the effective potential cannot, when calculated in the background
of the bounce, be expanded in a derivative expansion.  Consequently,
the imaginary part of the effective potential does not explicitly
enter the bubble nucleation calculation and the problems of
interpretation which it would entail are avoided.

\vskip 14truept

This work was supported in part by the U.~S.~Department of Energy.

\vskip 28truept
\noindent{\bf REFERENCES}
\vskip 14truept

\tenrm

\parindent=0.5truein

\noindent 1. S.~Coleman, \pr{15}{2929}77.

\noindent 2. S.~Coleman and E.~J.~Weinberg, \pr7{1888}73.

\noindent 3. See, e.g.,  P.~Frampton, \prl{37}{1378}76 and
Phys.~Rev.~D{\bf 15},\break \indent 2722 (1977); A.~Linde,
\pl{70}{306}77.

\noindent 4. E.~J.~Weinberg, \pr{47}{4614}93.

\noindent 5. C.~G.~Callan and S.~Coleman, \pr{16}{1762}77.

\noindent 6. E.~J.~Weinberg and A.~Wu, \pr{36}{2474}87.

\noindent 7. R.~Jackiw, \pr9{1686}74.

\noindent 8. N.~K.~Nielsen, \np{101}{173}75.

\noindent 9. R.~Fukuda and T.~Kugo, \pr{13}{3469}76.

\noindent 10. I.~J.~R.~Aitchison and C.~M.~Fraser, Ann. Phys. {\bf
156}, 1 (1984).

\vfill
\eject

\end